\documentclass[aps,prl,twocolumn]{revtex4}
\usepackage{amssymb}

\usepackage{graphicx}

\begin{document}

\title{Non-local Magnetic Field-tuned Quantum Criticality in Cubic CeIn$_{3-x}$Sn$_x$ ($x=$~0.25)}
\author{A.~V.~Silhanek,$^1$ Takao~Ebihara,$^2$, N.~Harrison,$^1$ M.~Jaime,$^1$, Koji~Tezuka$^2$, V.~Fanelli,$^1$, and C.~D.~Batista,$^1$ }
\affiliation{$^1$National High Magnetic Field Laboratory, Los Alamos National Laboratory,
MS E536, Los Alamos, NM 87545, USA\\
$^2$Department of Physics, Shizuoka University, Shizuoka 422-8529, Japan
}
\date{\today}

\begin{abstract}
We show that antiferromagnetism in lightly ($\approx$~8~\%) Sn-doped CeIn$_3$ terminates at a critical field $\mu_0H_{\rm c}=$~42~$\pm$~2~T. Electrical transport and thermodynamic measurements reveal that the effective mass $m^\ast$ does not diverge, suggesting that cubic CeIn$_3$ is representative of a critical spin-density wave (SDW) scenario, unlike the local quantum critical points reported in lower-symmetry systems such as CeCu$_{6-x}$Au$_x$ and YbRh$_2$Si$_{2-x}$Ge$_x$. The existence of a maximum in $m^\ast$ at a lower field $\mu_0H_{\rm x}=$~30~$\pm$~1~T may be interpreted as a field-induced crossover from local moment to SDW behavior as the magnitude of the antiferromagnetic order parameter falls below the Fermi bandwidth.
\end{abstract}

\pacs{PACS numbers: ..............................}
\maketitle

\begin{figure}[tbh]
\centering \includegraphics*[scale=0.9]{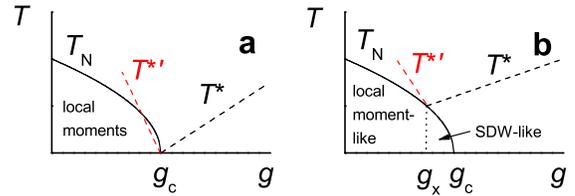}
\caption{Schematic of antiferromagnetic quantum criticality tuned at $g=g_{\rm c}$ according to (a) the locally critical scenario and (b) the SDW scenario, where $T$ represents the temperature scale respectively. The dotted line at $g_{\rm x}$ in (b) separates regions where the antiferromagnetism is predominantly local moment-like and SDW-like. The parameter $g$ can correspond to $p$, $H$ or $x$, depending on the system.}
\label{schematic}
\end{figure}

When the Neel temperature of an antiferromagnet is tuned to absolute zero at a quantum critical point (QCP), the uncertainty principle leads to a divergence in the characteristic lengthscale of the fluctuations of the staggered-moment order parameter $\Psi$~\cite{sachdev1}. In itinerant $d$- or $f$-electron antiferromagnets, strong on-site correlations often cause the renormalized Fermi bandwidth $k_{\rm B}T^\ast$ to become comparable to the Neel temperature $T_N$. A potential locally critical scenario arises in which the extent to which the $d$- or $f$-electrons {\it locally} contribute charge degrees of freedom to the Fermi liquid becomes subject to fluctuations at the QCP~\cite{coleman1,si1}.  Their effective localization is conditional upon the inequality $ T_N > T^\ast$ being satisfied~\cite{note0}, necessitating $T^\ast\rightarrow$~0 at the QCP as depicted in Fig.~\ref{schematic}a. Several $f$-electron antiferromagnets, including CeCu$_{6-x}$Au$_x$~\cite{stockert1,si1}, YbRh$_2$Si$_{2-x}$Ge$_x$~\cite{custers1} and CeRhIn$_5$~\cite{shishido1}, appear to provide examples of such behavior as function of pressure $p$, magnetic field $H$ or chemical substitution $x$. However, no experiment has yet been able to gauge the extent to which local criticality requires the spin fluctuations to be two-dimensional (2D)~\cite{coleman1,si1}. Were this an absolute requirement, the unambiguously three-dimensional (3D) spin fluctuation spectrum of cubic CeIn$_3$ should then provide the essential $f$-electron counterexample to local criticality~\cite{coleman1,lawrence1}. In such a case, one might expect a quantum critical spin-density wave (SDW) scenario to prevail~\cite{hertz1} in which $T^\ast$ remains finite at the QCP as depicted in Fig.~\ref{schematic}b.

In spite of CeIn$_3$ being one of the two original $f$-electron antiferromagnets in which non-Fermi liquid behavior and superconductivity were reported together under pressure~\cite{mathur1}, the comparatively large Neel temperature ($T_{\rm N}\approx$~10~K) requires rather extreme experimental conditions for its complete suppression; i.e. $p_{\rm c}\approx$~25~kbar and $\mu_0H_{\rm c}\approx$~61~T~\cite{ebihara1}. Neither the steep gradient $\partial T_{\rm N}/\partial p$ at large $p$~\cite{mathur1} nor the use of pulsed magnetic fields for $\mu_0H>$~45~T are amenable to precision tuning of the temperature dependent electrical resistivity $\rho(T)$ or direct measurements of the specific heat $C_p(T)$. Prior studies~\cite{mathur1, knebel1, kawasaki1} had therefore been unable to determine the applicable scheme in Fig.~\ref{schematic}.

In this paper, we utilize the fact that Sn-substitution of only $\sim$~8~\% of the In sites in CeIn$_3$ (yielding CeIn$_{2.75}$Sn$_{0.25}$) reduces $T_{\rm N}$ to $\approx$~6.4~K~\cite{pedrazzini1,note4} so as to enable quantum criticality of the same type II antiferromagnetic phase as in pure CeIn$_3$ to be tuned by static magnetic fields $\mu_0H\leq$~45~T.  This also enables us to avoid the technical difficulties associated with performing $C_p(T)$ and magnetization $M_z(H)$ measurements in combined high pressure, strong magnetic field conditions. Sn modifies the electronic structure by reducing the separation between the Fermi energy and the core $4f$-electron level, pushing this system further towards mixed valence~\cite{murani1}.  Prior measurements on single crystalline pure CeIn$_3$ had shown $T_{\rm N}(H) $ to be independent of the orientation of $H$, enabling the use of polycrystalline samples. Polycrystalline CeIn$_{3-x}$Sn$_x$ buttons with concentrations 0~$<x<$~0.75 are prepared by arc melting the appropriate quantities of 99.9, 99.999 and 99.9999~\% pure Ce, Sn and In respectively, with 5 further arc melts performed after flipping the button between melts for the purposes of homogenization. Samples cut from this button have $\rho(T)$ and $C_p(T)$ behaviors reproducing those obtained by Pedrazzini {\it et al.}~\cite{pedrazzini1}. In-situ $C_p(T)$ and $\rho(T)$ measurements on well characterized samples are then extended to fields $\mu_0H\leq$~45~T at temperatures $T\gtrsim$~1.5~K.

\begin{figure}[tbh]
\centering \includegraphics*[scale=1,angle=0]{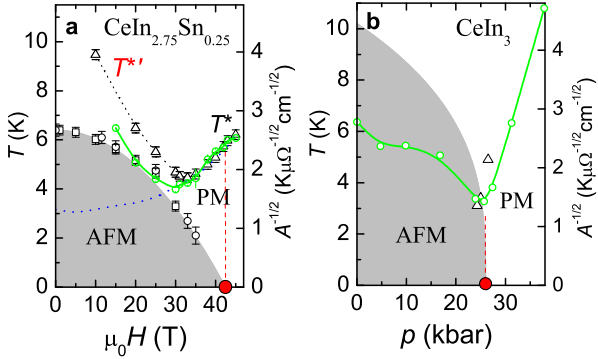}
\caption{(a) $T,H$ phase diagram of CeIn$_{2.75}$Sn$_{0.25}$ extracted from $C_p(T)$ ($\circ$ symbols) and $\rho(T)$ data ($\square$ symbols) as described in the text, with the grey region represents the aniterromagnetic (AFM) phase under the fitted $T_{\rm N}=T_{\rm N,0}(1-(H/H_{\rm c})^2)$ curve. $\triangle$ symbols delineate maxima in $\partial\rho/\partial T$ which are approximately representative of the Fermi bandwidth $T^\ast$ in the paramagnetic (PM) region. Green circles represent $A^{-1/2}$ in the low temperature limit $T\ll T^\ast$. The red dashed line represents $H_{\rm c}$, while the blue dotted line is $1/\chi^\prime=\partial H/\partial M$ approximately determined from $M_z(H)$ after smoothing and rescaling. All other lines are spline fits between data points. (b) The equivalent phase diagram versus $p$ instead of $H$ using the available data of Knebel {\it et al.}~\cite{knebel1}. Here the $\triangle$ symbols represent $T^\ast$ after Kawasaki et al.~\cite{kawasaki1}
}
\label{diagram}
\end{figure}

\begin{figure}[tbh]
\centering \includegraphics*[scale=1.2,angle=0]{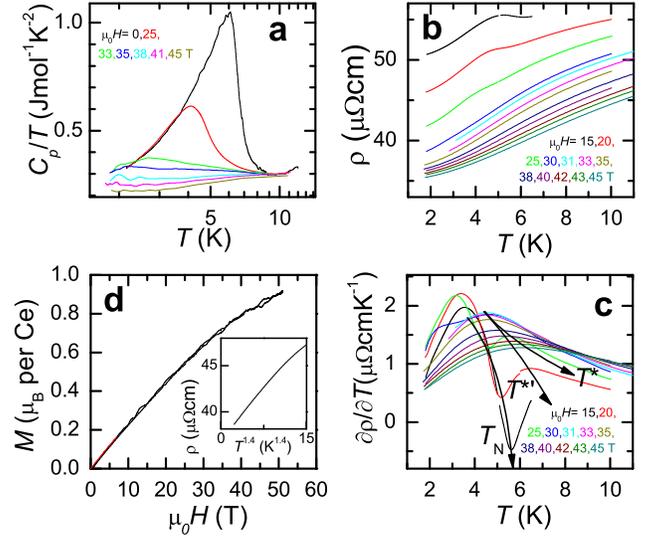}
\caption{(a) $C_p(T)/T$ raw data at selected magnetic fields $H$. (b) $\rho(T)$ data at selected magnetic fields. (c) The differential resistivity $\partial\rho(T)/\partial T$ obtained after polynomial smoothing the raw $\rho(T)$ data. (d) Magnetization of CeIn$_{2.75}$Sn$_{0.75}$ at 450~mK. The inset shows $\rho(T)$ at 30~T plotted versus $T^{1.4}$.}
\label{anomaly}
\end{figure}

Figure \ref{diagram} shows the $T,H$ phase diagram of CeIn$_{2.75}$Sn$_{0.25}$ extracted from the raw $C_p(T)/T$ and $\rho(T)$ data presented in Figs. \ref{anomaly}a and b respectively. The transition at $T_{\rm N}$ corresponds to a minimum in $\partial (C_p(T)/T)/\partial T$ ($\circ$ symbols), which can be further identified with a minimum in $\partial\rho(T)/\partial T$ in Fig.~\ref{anomaly}c ($\square$ symbols) for $\mu_0H\lesssim$~30~T. An empirical fit of $T_{\rm N}=T_{\rm N,0}(1-(H/H_{\rm c})^2)$ to the $\circ$ and $\square$ data points in Fig.~\ref{diagram}a yields  $\mu_0H_{\rm c}=$~42~$\pm$~2~T. The presence of an anomaly in $C_p(T)/T$ at 35~T and its absence at 45~T are consistent with the above estimate for $H_{\rm c}$. Several factors make the present data set consistent with the transition remaining of 2$^{\rm nd}$ order as $H\rightarrow H_{\rm c}$. These include the vanishing magnitude of the anomaly in $C_p(T)/T$ in Fig.~\ref{anomaly}a, the absence of a jump in the uniform magnetization $M_z(H)$ at 450~mK in Fig.~\ref{anomaly}d and the rather shallow slope $\partial T_{\rm N}/\partial H$ of the phase boundary in Fig.~\ref{diagram} such that $H_{\rm c}\partial T_{\rm N}/\partial H\ll p_{\rm c}\partial T_{\rm N}/\partial p$~\cite{ebihara1}.

The limiting value of $C_p(T)/T$ as $T\rightarrow$~0 provides an estimate of the coefficient $\gamma$ that accounts for the electronic contribution $\gamma T$ to $C_p(T)$. It is immediately apparent from the raw $C_p(T)/T$ data shown in Fig.~\ref{anomaly}a that, in ageement with the SDW scenario for a 3D system \cite{hertz1}, $\gamma$ does not exhibit any obvious signs of an emerging logarithmic divergence at or near $H_{\rm c}$. This is in stark contrast to the case of the $H$-tuned QCPs in CeCu$_{6-x}$Au$_x$~\cite{stockert1} and YbRh$_2$Si$_{2-x}$Ge$_x$~\cite{custers1}. The present findings are nevertheless consistent with the absence of a divergence in $m^\ast(H,p)$ obtained from $p$-and $H$-dependent dHvA experiments on pure CeIn$_3$ for ${\bf H}\|<100>$~\cite{ebihara1,endo1,note3}. 

\begin{figure}[tbh]
\centering \includegraphics*[scale=0.8]{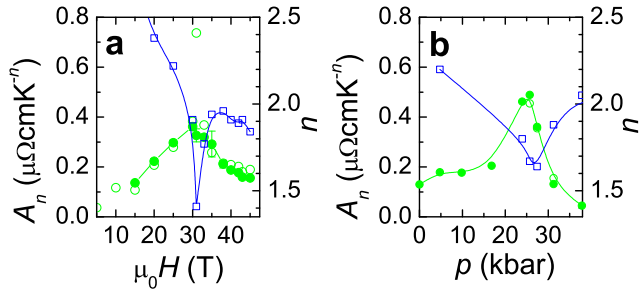}
\caption{Coefficient $A$ for CeIn$_{3-x}$Sn$_x$ estimated from $\partial\rho(T)/\partial T$. (a) shows $A$ (filled green circles) estimated as a function of $H$ for $x=$~0.25 and $n=$~2 with open circles showing estimates of $A_n$ where $n$ (open blue squares) is allowed to vary in order to facilitate a comparison with similar measurements under $p$. Spline fits are draw to quide the eye. (b) shows similar $A$ data for $x=$~0 as a function of $p$ obtained by Knebel {\it et al}.~\cite{knebel1}.}
\label{A}
\end{figure}

The electrical transport measurements presented in Fig.~\ref{anomaly}b further support the absence of a divergence at $H_{\rm c}$. Its derivative $\partial\rho(T)/\partial T$ shown in Fig.~\ref{anomaly}c yields a maximum at $T^\ast$ (or $T^{\ast\prime}$ for $\mu_0H<$~30~T) plotted in Fig.~\ref{diagram}a and an approximately linear region ($\approx 2AT$ where $A\propto m^{\ast 2}$) for $T\lesssim T^\ast$ over which the Fermi liquid behavior $\rho=\rho_0+AT^2$ appears to hold~\cite{kadowaki1} (a possible exception being at $\approx$~30~T;  see Fig.~\ref{anomaly}d inset and Fig.~\ref{A}a). On fitting $\rho=\rho_0+AT^2$ to these $T\lesssim T^\ast$ regions, $A(H>H_{\rm c})$ (solid circles) in Fig.~\ref{A}a can be seen to exhibit a qualitatively similar behavior to $A(p>p_{\rm c})$ in Fig.~\ref{A}b measured by Knebel {\it et al}~\cite{knebel1}. In order to understand how the Fermi liquid develops as $H\rightarrow H_{\rm c}$ and $p\rightarrow p_{\rm c}$, it is instructive to plot $A^{-1/2}(H,p)\propto 1/m^\ast$ in Fig.~\ref{diagram}. This can be seen to scale with $T^\ast(H)$ for $H>H_{\rm c}$, as observed within the paramagnetic regime of other materials~\cite{custers1,kim1} implying that $T^\ast$ approximately corresponds to the Fermi bandwidth. Both $A^{-1/2}(H)$ and $T^\ast(H)$ also scale with $1/\chi^\prime(H)$ for $H>H_{\rm c}$ in Fig.~\ref{diagram}a, indicating that both the electrical transport and magnetization are consistent with a Fermi liquid comprised of partially polarized quasiparticle bands. Were $m^\ast(H,p)$, $\gamma(H,p)$ and $A(H,p)$ actually to diverge, we would expect $A(H,p)^{-1/2}$ to vanish at $H_{\rm c}$ and $p_{\rm c}$~\cite{coleman1}. Instead, $A^{-1/2}\propto T^\ast$ can be seen to vary in an approximately linear fashion with $H$ and $p$, intercepting $H=H_{\rm c}$ and $p=p_{\rm c}$ at a finite value in both Figs.~\ref{diagram}a and b.

Taken together, these findings imply that the quantum fluctuations of $\Psi$ have a similarly weak effect on the Fermi liquid properties at both $p_{\rm c}$ in pure CeIn$_3$ and $H_{\rm c}$ in CeIn$_{2.75}$Sn$_{0.25}$. Hence, while CeCu$_{6-x}$Au$_x$~\cite{si1}, YbRh$_2$Si$_{2-x}$Ge$_x$~\cite{custers1} and CeRhIn$_5$~\cite{shishido1} may be considered consistent with the local criticality picture in which the Fermi surface topology undergoes a dramatic change at the QCP~\cite{si1}, this clearly cannot be the case for CeIn$_3$ and CeIn$_{2.75}$Sn$_{0.25}$.

One unique advantage of the present $H$-tuned QCP is that the slope of the phase boundary as $T_{\rm N}\rightarrow$~0 in Fig.~\ref{diagram} is sufficiently gradual as a function of $H$ that we can continue to observe the proportionality between $A^{-1/2}$, $T^\ast$ and $\chi^{\prime -1}$ over a significant $\sim$~12~T wide interval in field below $H_{\rm c}$. This continues until $A^{-1/2}(H)$ (rescaled in units of kelvin) intersects $T_{\rm N}(H)$, whereupon they diverge. The coincidence of the minimum in $A(H)^{-1/2}$ (or maximum in $A$) at $\mu_0H_{\rm x}=$~30~$\pm$~1~T with the point on the phase diagram at which $T_{\rm N}\approx T^\ast$ in Fig.~\ref{schematic}b is strongly suggestive of its association with a crossover from SDW-like behavior to local moment-like behavior of the form depicted in Fig.~\ref{schematic}b. Such a crossover would directly affect the degree to which the $f$-electrons contribute charge degrees of freedom to the Fermi liquid~\cite{note0}. 

Outside the antiferromagnetic phase, the $f$-electrons are hybridized with the conduction electrons giving rise to heavy Fermi liquid composed of renormalised quasiparticle bands that incorporate the $f$-electron charge degrees of freedom~\cite{hewson1}. The on-site correlations weaken with increasing $g$ (i.e. $p$ or $H$), causing $A^{-1/2}$, $T^\ast$ and $\chi^{\prime -1}$ {\it all} to increase. Provided $T_N \ll T^\ast$, a weak coupling SDW that gaps parts of the Fermi surface can form~\cite{coleman1,hertz1} with no significant change in the overall effect of the correlations on $T^\ast$. To compute of the influence of a finite antiferromagnetic order parameter $\Psi$ on the quasiparticle bands, it is necessary to combine the effect of the on-site Kondo 
interaction with the ${\bf K}={\bf Q}$ (${\bf Q}$ is the antiferromagnetic wave vector) Bragg scattering produced by the underlying antiferromagnetic
structure. 

At the local QCP depicted in Fig.~\ref{schematic}a, $T^\ast$ vanishes precisely at $g=g_{\rm c}$ because the Kondo
screening is supressed by the magnetic interactions. The Fermi surface undergoes a large reconstruction accross the 
QCP since the development of antiferromagnetism inhibits the $f$-electrons from contributing charge degrees of freedom to the Fermi liquid~\cite{coleman1,si1}, possibly leading to the emergence of new Fermi liquid with a different (smaller) Fermi surface topology and a different characteristic $T^{\ast\prime}$ that now increases with decreasing $g$~\cite{custers1} (depicted in red in Fig.~\ref{schematic}). In the quantum critical SDW scenario depicted in Fig.~\ref{schematic}b, however,  the Kondo screening is 
not supressed at the QCP and the Fermi surface evolves smoothly across the phase transition. Should $H_{\rm x}$ correspond to $g_{\rm x}$ in Fig.~\ref{schematic}b, the loss in proportionality between $A^{-1/2}$ and $T^{\ast\prime}$ for $H<H_{\rm x}$ can be explained as a consequence of the former being determined for $T\ll T_{\rm N}$ and the latter for $T>T_{\rm N}$, with the Fermi surface topology being modified by $\Psi$ only in the former case. Note that $\chi^{\prime -1}$ departs from $A^{-1/2}$ more strongly once $H<H_{\rm c}$, because it now includes the contribution to the susceptibility from the canted ordered moments.

Our ability to infer $H_{\rm x}\neq H_{\rm c}$ in the present study stems from the fact that $H_{\rm c}\partial T_{\rm N}/\partial H\ll p_{\rm c}\partial T_{\rm N}/\partial p$ close to the QCP. More finely $p$-tuned NQR studies~\cite{kawasaki1} have recently shown that the lowest Fermi liquid temperature (equivalent to $T^\ast$) occurs at a pressure that is $\approx$~0.75~kbar lower than $p_{\rm c}$, suggesting also that $p_{\rm x}\neq p_{\rm c}$. This observation combined with the similar values of $A^{-1/2}$ at which $A^{-1/2}(H,p)\propto T^\ast(H,p)$ intercepts $T_{\rm N}$ and the similar minimum $\sim$~1.5 value of the exponent $n$ (blue squares) obtained on fitting $\rho=\rho_0+A_nT^n$ to the electrical resistivity data in Fig.~\ref{A} suggests that the $p$- and $H$-dependent QCP's are related. Both $p$ and $H$ lead to a gradual reduction in the size of the staggered moment and monotonic increase in $T^\ast$ as the effect of the on-site correlations is suppressed. In the former case this is caused by an increase overlap between the $f$-electron orbitals while in the latter case it is caused by the progressive polarization of the $f$-electrons by the Zeeman interaction.

The monotonic $M_z(H)$ of CeIn$_{3-x}$Sn$_x$ is yet another quality that can be attributed to its cubic symmetry. Magnetically anisotropic systems, by contrast, sometimes undergo an abrupt increase in the uniform magnetization (i.e. metamagnetism) at a characteristic magnetic field $H_{\rm m}$ that is unrelated to antiferromagnetism~\cite{flouquet1}. Systems that combine both antiferromagnetism and metamagnetism tend to exhibit a more complicated field-dependent behavior. CeRh$_2$Si$_2$~\cite{hamamoto1} and UPd$_2$Al$_3$~\cite{terashima1} are two examples of systems in which metamagnetism causes antiferromagnetism to terminate prematurely at $H_{\rm m}$ owing to the fact that $M_z$ and $\Psi$ compete for the same spin degrees of freedom.

In summary, we find that the light Sn-doping of CeIn$_3$ facilitates observation of a $H$-tuned QCP at $H_{\rm c}=$~42~$\pm$~2~T, enabling $\rho(T)$ and $C_p(T)$ measurements to be performed in static magnetic fields $\mu_0H<$~45~T. Neither $A^{-1/2}$, $T^\ast$ nor $\chi^{\prime -1}$ collapse to zero at the QCP, indicating that the system continues to exhibit conventional Fermi liquid behavior as $\Psi\rightarrow$~0, suggestive of a 3D quantum critical SDW scenario as opposed to a locally critical scenario. We attribute the observation of a minimum in $A^{-1/2}$ at a somewhat lower field $H_{\rm x}=$~30~$\pm$~1~T to the fact that the Kondo screening is not supressed at the 
QCP. Correspondingly, $T^{\ast}$ continues decreasing until it crosses the antiferromagnetic phase boundary ($T^{\ast}=T_N$).
At that point, $T^{\ast}$, now $T^{\ast\prime}$, starts to increase because the order parameter $\Psi$ becomes stonger when the magnetic field is reduced   Fig.~\ref{schematic}b. A bigger value of $\Psi$ leads to a weaker effective Kondo coupling between the $f$- and the conduction 
electrons. Similarities in the electrical transport behavior of CeIn$_{2.75}$Sn$_{0.25}$ at $H_{\rm x}<H_{\rm c}$ and CeIn$_3$ at $p_{\rm x}<p_{\rm c}$~\cite{kawasaki1} suggests that a similar QCP may be accessed in both cases. 

The possibility of universality in the $H$- and $p$-induced behavior warrants further investigations of pure CeIn$_3$ by means of electrical transport measurements under combined high $H$ and $p$ conditions. Realization of a situation in which the maximum in $m^\ast$ and antiferromagnetic QCP occur at distinctly different pressures in CeIn$_3$ (i.e.$p_{\rm x}\neq p_{\rm c}$ in Fig.~\ref{schematic}b)~\cite{kawasaki1} may provide a unique opportunity to identify the key factors required to optimize unconventional supercnductivity~\cite{mathur1}. 

This work was performed under the auspices of the National Science
Foundation, the Department of Energy (US) and the State of Florida. TE would like  to thank the Suzuki Foundation for its support in travel expenses, the Casio Science Promotion foundation for additional financial support and Dr. H. Kitazawa of NIMS, Japan for utilizing his arc furnace. AVS acknowledges Pablo Pedrazzini for useful discussions while VF would like to thank the NHMFL In-House Research Program.


\begin{thebibliography}{99}
\bibitem{note0} In the local quantum critical picture, the magnetic interactions supress the Kondo screening 
and, correspondingly, the coherence temperature $T^\ast$ vanishes at the QCP. Since the RKKY mediated antiferromagnetic order competes with the Kondo screening responsible for formation of the heavy Fermi liquid, it is unlikely that a heavy Fermi liquid that includes the $f$-electron charge degrees of freedom can continue to survive inside the antiferromagnetic phase. Decoupling of the $f$-electron charge degrees of freedom from the Fermi liquid should eventually lead to local moment ordering and development of a weakly correlated Fermi liquid with a higher Fermi temperature.
\bibitem{note3} $H$ orientation-dependent dHvA experiments performed on pure CeIn$_3$ at ambient pressure inside the antiferromagnetic phase indicate that $m^\ast$ increases as $H\rightarrow H_{\rm c}$ at `hot spots' that occupy $\lesssim$~2~\% of the sperical d-sheet Fermi surface area~\cite{ebihara1}, possibly contributing to the $H$- and $p$-dependences of $A$ inside the AFM phase in Figs.~\ref{diagram} a and b.
\end{thebibliography}
\end{document}